\documentclass[prd,twocolumn,showpacs,amssymb]{revtex4}

\usepackage{graphicx}

\begin{document}

\title{Nonstrange hybrid compact stars with color superconducting matter}

\author{Igor Shovkovy}
  \email{shovkovy@th.physik.uni-frankfurt.de}
  \altaffiliation[On leave
of absence from ]{%
       Bogolyubov Institute for Theoretical Physics, 
       03143, Kiev, Ukraine
}%

\author{Matthias Hanauske}
  \email{hanauske@th.physik.uni-frankfurt.de}

\author{Mei Huang}
  \email{huang@th.physik.uni-frankfurt.de}
  \altaffiliation[Also at ]{%
       Physics Department, Tsinghua University,
       Beijing 100084, China
}%
\affiliation{%
       Institut f\"{u}r Theoretische Physik, 
       J.W. Goethe-Universit\"{a}t, 
       D-60054 Frankurt/Main, Germany
}%

\date{\today}

\begin{abstract}
Realistic nonstrange hybrid compact stars with color superconducting
quark matter in their interior are constructed. It is shown that a
positively charged two-flavor color superconducting phase could naturally
appear in the core of a hybrid star as one of the components of a
globally neutral mixed phase. The negatively charged normal quark phase
is the other component of the mixed phase. The quark core of the star is
surrounded by another mixed phase made of hadronic and normal quark
matter. The two mixed phases are separated by a sharp interface. Finally,
the lowest density regions of the star are made of pure hadronic matter
and nuclear crust.
\end{abstract}



\maketitle

\section{Introduction}

At large densities, quantum chromodynamics (QCD) becomes a weakly
interacting theory of quarks \cite{ColPer}. The quarks tend to form a
highly degenerate Fermi surface. Because of the existence of an
attractive interaction in the color-antitriplet channel, the famous
Cooper instability develops, and the ground state of the system is
rearranged \cite{old,bl,cs}. The rearrangement is similar to the
Bardeen-Cooper-Schrieffer (BCS) mechanism of low temperature
superconductivity in metals and alloys \cite{BCS}. The ground state of
dense quark matter is a color superconductor. It is characterised by
spontaneous breaking (through the Anderson-Higgs mechanism) of the
non-Abelian SU(3)$_c$ color gauge group rather than the Abelian
U(1)$_{em}$ group of electromagnetism.
 
Weakly interacting quark matter at asymptotic densities was studied from
first principles in Refs. \cite{weak,pr2,weak-cfl}. These studies proved
that quark matter is indeed a color superconductor at sufficiently high
densities. An explicit asymptotic expression for the superconducting
order parameter was also derived in Refs. \cite{weak,pr2,weak-cfl,drqw}.
However, the microscopic studies are not very reliable quantitatively at
realistic densities that exist in nature (e.g., inside compact stars).
The corresponding running coupling constant of QCD defined at the
relevant scale of the quark chemical potential is large, and therefore
the results cannot be trusted.

It might be appropriate to mention that the creation of cold dense quark
matter with color superconducting properties in heavy ion collisions
seems very unlikely to us (see, however, some speculations in Ref.
\cite{kitazawa}). In particular, it does not look plausible that one
could avoid producing a large entropy per baryon in any type of a heavy
ion collision. Therefore, a cold and dense environment, needed to support
color superconductivity, could hardly be formed. A typical central region
of a compact star, on the other hand, appears to be a very natural place
where sufficiently cold and dense matter could exist. Indeed, the
estimated central densities of such stars might be as large as $10
\rho_{0}$ (where $\rho_{0} \approx 0.15$ fm$^{-3}$ is the saturation
density), while their temperatures are in the range of $10$ to $100$ keV.

In order to study quark matter at realistic densities, currently there is
no other real alternative but to rely on various phenomenological models
(such, for example, as Nambu-Jona-Lasinio type models
\cite{cs,4fermi,cfl,unlock}). The obvious shortcoming of this approach is
the same as that of the microscopic approach: the results should always
be interpreted with caution and treated as qualitative rather than
quantitative. Nevertheless, if one accepts the possibility of color
superconductivity in quark matter at realistic densities, existing in
compact stars, it is not so important which specific model is used to
estimate the effect of color superconductivity on properties of compact
stars.

In general, QCD at high baryon density has a very rich phase structure.
There are many possible color superconducting phases of quark matter
made of one, two and three lightest quark flavors. Each of them is
characterised by a unique symmetry breaking pattern and by a specific
number of bosonic as well as fermionic gapless modes. In the rest of this
paper, we are going to concentrate our attention almost exclusively on
matter with two quark flavors. In order to better understand our
motivation, let us also briefly mention other possibilities.

A one-flavor color superconductor is characterised by a condensate of
Cooper pairs made of the same quark flavor. Thus, the corresponding wave
function is symmetric in flavor. By taking into account that the
condensation results from the attractive interaction between quarks in the
color antitriplet antisymmetric channel, one finds that a spin zero
configuration is forbidden by the Pauli principle. Therefore, it is a much
weaker spin-1 condensate that gives rise to color superconductivity in
one-flavor dense quark matter \cite{bl,pr2,spin-1}. The recent study in
Ref. \cite{swr-meissner} suggests that the presence of such a phase may
still affect some properties of compact stars.

Perhaps, the most interesting color superconducting phase appears in
quark matter with three light flavors. It is the so called
color-flavor-locked (CFL) phase \cite{cfl}. In this phase the original
gauge symmetry SU(3)$_{c}$ and the global chiral symmetry SU(3)$_{L}
\times$ SU(3)$_{R}$ break down to a global diagonal $SU(3)_{c+L+R}$
subgroup. As a result, the quark quasiparticles get large gaps in their
energy spectra and decouple from the low energy dynamics. Because of the
Higgs mechanism all gluons also become massive and decouple. Then, it
appears that the low energy spectrum of the CFL phase is dominated by one
Nambu-Goldstone boson and nine massive pseudo-Nambu-Goldstone bosons
\cite{cfl}. This observation has interesting implications for the
transport properties of the CFL phase \cite{SE}.

In order to have the CFL phase in compact stars, the strange quark should
be sufficiently light at the corresponding densities. However, the actual
density dependence of the constituent, medium modified mass of the
strange quark is not known in QCD. Taking this into account, one could
imagine that the strange quark might be too heavy even to appear in
baryonic matter at realistic conditions inside compact stars. Then, of
course, the CFL quark phase could not be realized. In yet another
scenario, the strange quarks could appear inside stars, but they would
not participate in Cooper pairing. Then, a mixture of 2SC matter and
normal matter of strange quarks could form. We denote such a phase
2SC+s.

Matter in the bulk of a compact star should be neutral (at least, on
average) with respect to electric as well as color charges. Otherwise,
the gravitational force would not be able to hold the star together. Also
such matter should remain in $\beta$-equilibrium. These conditions turn
out to play an important role in determining which color superconducting
phases can and which cannot exist inside stars. Indeed, the strongest
color superconducting phases appear when the Fermi momenta of different
quark flavors are approximately equal. By enforcing the neutrality and
$\beta$-equilibrium conditions, however, one does not necessarily get the
Fermi surfaces of quarks that are best suitable for Cooper pairing. In
some cases, these conditions may even appear to be inconsistent with
color superconductivity.

Recently, it was argued in Ref. \cite{absence2sc} that imposing the
charge neutrality condition on quark matter inside compact stars would
prevent the appearance of the 2SC+s phase, and would favor the appearance
of the CFL phase instead. (Note that the strange quark mass was chosen
too small to allow the appearance of the pure 2SC phase in Ref.
\cite{absence2sc}.) This conclusion was essentially confirmed in a more
rigorous study of Ref. \cite{neutral_steiner} where it was found that the
2SC+s phase could exist only in a narrow window (about $10$ or $15$ MeV
wide) of the baryon chemical potential around the midpoint $\mu_{B}/3
\approx 450$ MeV. In this phase, a positive charge of the 2SC condensate
is compensated by a negative charge of unpaired strange quarks. At
smaller values of the chemical potential, no conventional 2SC was
detected in Ref. \cite{neutral_steiner}. It was shown in Ref. \cite{HS}
that a new type of neutral 2SC phase, the so called gapless 2SC phase
develops at the corresponding densities. In this gapless 2SC phase, the
conventional relations between the number densities of the pairing quarks
are not valid.

While matter in the bulk of a star should satisfy the condition of charge
neutrality, this condition should not necessarily be enforced {\em
locally}. In fact, it is sufficient to make matter neutral on average, or
{\em globally}. This kind of matter appears naturally in mixed phases.
Recently, studies of various mixed phases with color superconducting
matter were presented in Ref. \cite{neutral_buballa}. In this paper, we
use this general idea to construct a realistic model of a compact star
with two-flavor color superconducting matter at the core (for
constructions using 2SC quark matter without mixed phases see Ref.
\cite{Blaschke_2sc,Ruester}, and for constructions using CFL quark matter
see Ref. \cite{super-dense,compact_CFL}).

This paper is organized as follows. In the next section, we present a
quark model with a four-fermion interaction that we use to derive the
quark matter equation of state. There, we also briefly discuss the phase
diagram of the model, and the subtleties of enforcing the electrical
neutrality condition in color superconducting matter. In Sec.
\ref{quark-Gibbs}, we present the Gibbs construction for the mixed phase
of normal and color-superconducting quark matter. In Sec.
\ref{low-density}, we discuss a model for hadronic matter, and the
corresponding equation of state with the charge neutrality condition
imposed. In the same section, we apply the Gibbs construction to derive
the equation of state of the hadron-quark mixed phase. The star
properties are discussed in Sec. \ref{star-properties}, and the summary
and outlook are given in Sec. \ref{summary}.

\section{Quark model}
\label{quark-model}

We start our analysis from discussing the quark model that we use to
derive the equation of state of baryonic matter at high density.  In this
study, we accept a rather conservative point of view that the strange
quark is sufficiently heavy and does not appear at baryon densities that
can be reached in compact stars without causing a gravitational
instability. It is also appropriate to mention that we aim at a general
qualitative rather than quantitative description of hybrid stars with
color superconducting matter in their interior. These assumptions justify
the use of the simplest SU(2) Nambu-Jona-Lasinio model of Ref.
\cite{huang_2sc} in our study. The explicit form of the Lagrangian
density reads:
\begin{eqnarray} 
{\cal L} & =
&\bar{q}(i\gamma^{\mu}\partial_{\mu}-m_0)q +
 G_S\left[(\bar{q}q)^2 + (\bar{q}i\gamma_5{\bf \vec{\tau}}q)^2\right]
\nonumber \\
 &+& G_D\left[(i \bar{q}^C  \varepsilon  \epsilon^{b} \gamma_5 q )
   (i \bar{q} \varepsilon \epsilon^{b} \gamma_5 q^C)\right],
\label{lagr}
\end{eqnarray}
where $q^C=C {\bar q}^T$ is the charge-conjugate spinor and $C=i\gamma^2
\gamma^0$ is the charge conjugation matrix. The quark field $q \equiv
q_{i\alpha}$ is a four-component Dirac spinor that carries flavor
($i=1,2$) and color ($\alpha=1,2,3$) indices.  ${\vec \tau} =(\tau^1,
\tau^2, \tau^3)$ are Pauli matrices in the flavor space, and
$(\varepsilon)^{ik} \equiv \varepsilon^{ik}$, $(\epsilon^b)^{\alpha
\beta} \equiv \epsilon^{\alpha \beta b}$ are antisymmetric tensors in 
flavor and color, respectively. We also introduced two independent
coupling constants in the scalar quark-antiquark and scalar diquark
channels, $G_S$ and $G_D$. The definition of this non-renormalizable NJL
model is complete after introducing a momentum cut-off $\Lambda$. 

The values of the parameters in the NJL model are the same as in Refs.
\cite{HS,huang_2sc}: $G_S=5.0163$ GeV$^{-2}$ and $\Lambda=0.6533$ GeV.
Here we also consider only the chiral limit with $m_0=0$. As for the
strength of the diquark coupling $G_D$, its value is taken to be
proportional to the quark-antiquark coupling constant, i.e., $G_D = \eta
G_S$ with $\eta=0.75$. The choice $\eta=0.75$ corresponds to the regime
of intermediate strength of the diquark coupling \cite{HS}. Regarding
this point, we mention that our final results would remain qualitatively
the same also for a wide range of weak and strong diquark couplings.

In $\beta$-equilibrium, the diagonal matrix of quark chemical potentials
is given in terms of baryonic, electric and color chemical potentials.
In particular,
\begin{equation}
\mu_{ij, \alpha\beta}= (\mu \delta_{ij}- \mu_e Q_{ij})
\delta_{\alpha\beta} + \frac{2}{\sqrt{3}}\mu_{8} \delta_{ij}
(T_{8})_{\alpha \beta},
\end{equation}
where $Q$ and $T_8$ are generators of U(1)$_{em}$ of electromagnetism and 
U(1)$_{8}$ subgroup of the color gauge group. The explicit expressions
for the quark chemical potentials read
\begin{eqnarray}
\mu_{ur} =\mu_{ug} =\mu -\frac{2}{3}\mu_{e} +\frac{1}{3}\mu_{8}, \\
\mu_{dr} =\mu_{dg} =\mu +\frac{1}{3}\mu_{e} +\frac{1}{3}\mu_{8}, \\
\mu_{ub} =\mu -\frac{2}{3}\mu_{e} -\frac{2}{3}\mu_{8}, \\
\mu_{db} =\mu +\frac{1}{3}\mu_{e} -\frac{2}{3}\mu_{8}.
\end{eqnarray}
One should notice that, in general, there are two generators ($T_3$ and
$T_8$) in the center of the SU(3)$_{c}$ color group. Therefore, one could
introduce two chemical potentials for two different color charges.
However, we require that quark matter in the 2SC ground state remain
invariant under the SU(2)$_{c}$ color gauge subgroup. This condition
forbids the introduction of the second nontrivial color chemical
potential $\mu_{3}$.

The effective potential for quark matter at zero temperature and in
$\beta$-equilibrium with electrons takes the form \cite{HS}
\begin{equation}
\label{potential}
\Omega = \Omega_{0}-\frac{\mu_e^4}{12 \pi^2}
+\frac{m^2}{4G_S}+\frac{\Delta^2}{4G_D} 
- \sum_{a} \int\frac{d^3 p}{(2\pi)^3} |E_{a}|,
\end{equation}
where $\Omega_{0}$ is a constant added to make the pressure of the vacuum
zero. The sum in Eq. (\ref{potential}) runs over all (6 quark and 6
antiquark) quasiparticles. The dispersion relations and the degeneracy
factors of the quasiparticles read
\begin{eqnarray} 
E_{ub}^{\pm} &=& E(p) \pm \mu_{ub} , \hspace{26.6mm} [\times 1]
\label{disp-ub} \\
E_{db}^{\pm} &=& E(p) \pm \mu_{db} , \hspace{26.8mm} [\times 1]
\label{disp-db}\\
E_{\Delta^{\pm}}^{\pm} &=& \sqrt{[E(p) \pm \bar{\mu}]^2
+\Delta^2} \pm  \delta \mu .\hspace{6.2mm} [\times 2]
\label{2-degenerate}
\end{eqnarray}
Here we used the following shorthand notation:
\begin{eqnarray}
E(p) &\equiv& \sqrt{{\bf p}^2+m^2}, \\
\bar{\mu} &\equiv& \frac{\mu_{ur} +\mu_{dg}}{2} =
\frac{\mu_{ug}+\mu_{dr}}{2}
=\mu-\frac{\mu_{e}}{6}+\frac{\mu_{8}}{3}, \\
\delta\mu &\equiv& \frac{\mu_{dg}-\mu_{ur}}{2}
=\frac{\mu_{dr}-\mu_{ug}}{2}
=\frac{\mu_{e}}{2}.
\end{eqnarray}
After making use of the explicit quasiparticle dispersion relations, we
arrive at the following potential:
\begin{widetext}
\begin{eqnarray}
\Omega &=& \Omega_{0}-\frac{\mu_e^4}{12 \pi^2}
+\frac{m^2}{4G_S}+\frac{\Delta^2}{4G_D} - 2\int_{0}^{\Lambda} 
\frac{p^2 d p}{\pi^2} \Big(E(p)
+\sqrt{[E(p)+ \bar{\mu}]^2+\Delta^2}
+\sqrt{[E(p)-\bar{\mu}]^2+\Delta^2}\Big) \nonumber\\
&-&\int_{0}^{p^{(ub)}_{F}} \frac{p^2 d p}{\pi^2}[\mu_{ub}-E(p)]
- \int_{0}^{p^{(db)}_{F}} \frac{p^2 d p}{\pi^2}[\mu_{db}-E(p)]
- 2\theta \left(\delta\mu-\Delta\right)
\int_{\mu^{-}}^{\mu^{+}}
\frac{p^2 d p}{\pi^2}\Big(\delta\mu
- \sqrt{[E(p)-\bar{\mu}]^2+\Delta^2}\Big),
\label{pot-2sc}
\end{eqnarray}
\end{widetext}
where $\mu^{\pm}\equiv \bar{\mu}\pm \sqrt{(\delta\mu)^2-\Delta^2}$. 
Note that the physical thermodynamic potential that determines the
pressure, $\Omega_{\rm phys} =-P$, is obtained from $\Omega$ in
Eq. (\ref{pot-2sc}) after substituting the order parameter $\Delta$ that
solves the following gap equation:
\begin{equation}
\frac{\partial \Omega}{\partial \Delta} =0.
\label{gap-eq}
\end{equation} 
In addition, homogeneous quark matter in the bulk of a compact star
should be neutral with respect to color and electric charges. Therefore,
for such matter, one should also impose extra two (local) conditions:
\begin{eqnarray}
n_{8} &\equiv&
\frac{\partial \Omega}{\partial \mu_{8}}=0,
\label{color-neut}\\
n_{e} &\equiv&
\frac{\partial \Omega}{\partial \mu_{e}}=0.
\label{electr-neut}
\end{eqnarray}
The solution that satisfies the gap equation and both neutrality
conditions was studied in detail in Ref. \cite{HS}. It was found that the
corresponding phase of matter is the gapless 2SC phase. This phase has
the same symmetry of the ground state as the conventional 2SC phase. In
the low-energy spectrum, however, it has two additional gapless fermionic
quasiparticles.

In order to determine the most favorable phase of quark matter inside a
hybrid compact star, one has to consider all possible neutral phases, 
and compare their values of the pressure at a given baryon chemical
potential. The simplest candidates are the gapless 2SC and normal quark
phases. Both of these can be made locally neutral. For any given value of
the baryon chemical potential, gapless 2SC matter has a slightly higher
pressure than normal quark matter \cite{HS}. Numerically, we find that
the pressure difference $\delta P$ lies somewhere between 1 and 2.5
MeV/fm$^{3}$. Therefore, the gapless 2SC phase is slightly more favorable
than the neutral normal quark phase.

In addition to homogeneous (one-component) phases, one could also study
various mixed phases of two-flavor quark matter. The most promising
components for constructing mixed phases are (i) normal quark matter, (ii)
gapless 2SC matter, and (iii) ordinary 2SC matter. The first two of them
are included because they allow locally neutral phases by themselves. The
last one, on the other hand, is included because it is the strongest color
superconducting phase known. Clearly, this list of phases is not complete.
One could also take into consideration other phases, e.g., various
combinations of color superconductors with spin-1 condensates
\cite{pr2,spin-1,swr-meissner}. Our experience suggests that these latter
would lead to less favorable one- and multi-component constructions. In
order to prove this rigorously, a detailed study is required. This,
however, is outside the scope of the present paper.

Inside mixed phases, the charge neutrality is satisfied ``on average"
rather than locally. This means that different components of mixed phases
may have non-zero densities of conserved charges, but the total charge of
all components still vanishes. In this case, one says that the local
charge neutrality condition is replaced by a global one. Now, in absence
of the {\em local} neutrality, the pressure of each phase could be
considered as a function of baryon chemical potential, as well as a
function of chemical potentials related to other conserved charges (e.g.,
$\mu_{e}$ and $\mu_{8}$ in the case of two-flavor quark matter at hand).
In order to decide which mixed phase is the most favorable, one should
utilize the so-called Gibbs construction. The details of this
construction will be given in the next section.

While we intend to relax the local neutrality condition with respect to
the electric charge, in this paper we always enforce the condition of
local {\em color} neutrality. We cannot fully justify this constrain in
all situations where mixed phases appear. However, one could speculate
that color charged components are less likely to make energetically
favorable mixed phases because of strong color Coulomb forces. Besides, in
the two main constructions that we build in this paper, at least one of
the components (normal quark or hadronic matter) of a mixed phase is
expected to be locally color neutral. Therefore, our consideration here
remains rather general.

\begin{figure}
\includegraphics[bbllx=88,bblly=4,bburx=561,bbury=504,width=8cm]
{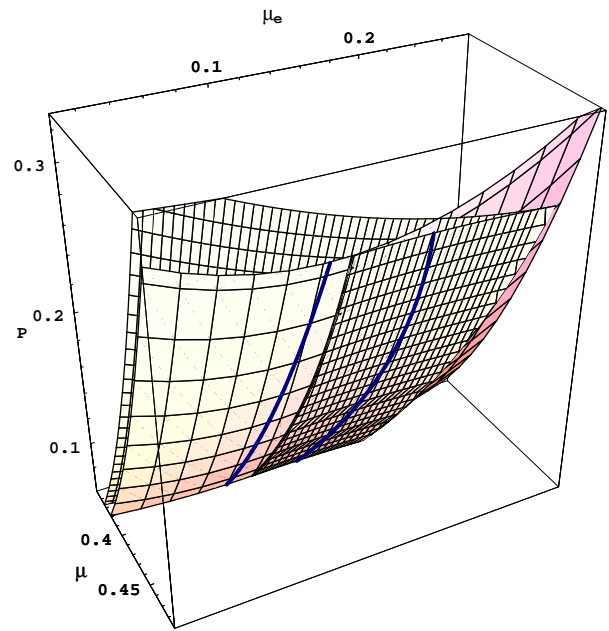}
\caption{\label{fig-back}
Pressure as a function of $\mu\equiv\mu_B/3$ and $\mu_e$ for the normal
and color superconducting quark phases. The dark solid lines represent
two locally neutral phases: (i) the neutral normal quark phase on the
left, and (ii) the neutral gapless 2SC phase on the right. The appearance
of the swallowtail structure is related to the first order type of the
phase transition in quark matter.}
\end{figure}

The pressure of the main three phases of two-flavor quark matter as a
function of the baryon and electric chemical potentials is shown in Fig.
\ref{fig-back}. In this figure, we also show the pressure of the neutral
normal quark and gapless 2SC phases (two dark solid lines). In full
agreement with the study of Ref. \cite{HS}, the surface of the gapless
2SC phase extends only over a finite range of the values of $\mu_{e}$. It
merges with the pressure surfaces of the normal quark phase (on the left)
and with the ordinary 2SC phase (on the right).

It is interesting to notice that the three pressure surfaces in Fig.  
\ref{fig-back} form a characteristic swallowtail structure. As one
could see, the appearance of this structure is directly related to the
fact that the phase transition between color superconducting and 
normal quark matter, which is driven by changing parameter $\mu_{e}$, is
of first order. In fact, one should expect the appearance of a similar
swallowtail structure also in a self-consistent description of the
hadron-quark phase transition. Such a description, however, is not
available yet.

>From Fig. \ref{fig-back}, one could see that the surfaces of normal  
and 2SC quark matter intersect along a common line. This means that the      
two phases have the same pressure along this line, and therefore could
potentially co-exist. Moreover, as is easy to check, normal quark
matter is negatively charged, while 2SC quark matter is positively
charged on this line.  This observation suggests that the appearance of   
the corresponding mixed phase is almost inevitable. The details of the
constructions are given in the next section.

\section{Gibbs construction within quark model}
\label{quark-Gibbs}

In this section we study the possibility of mixed phases in two-flavor
quark matter. As we have already mentioned in the previous section, we
restrict our consideration to only three most promising components: (i)
normal quark matter, (ii) gapless 2SC matter, and (iii) ordinary 2SC
matter. Before proceeding further, it would be appropriate to mention
that some mixed phase with color superconducting components were also
constructed in Ref. \cite{neutral_buballa}.

\begin{figure}
\includegraphics[bbllx=88,bblly=4,bburx=502,bbury=504,width=8cm]
{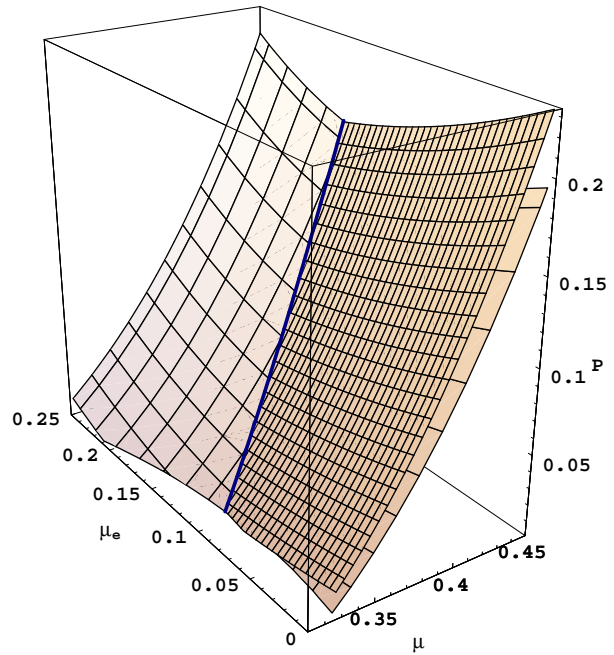}
\caption{\label{fig-front-quark}
Pressure as a function of $\mu\equiv\mu_B/3$ and $\mu_e$ for the normal
and color superconducting quark phases (the same as in 
Fig. \ref{fig-back}, but from a different viewpoint). The dark solid line
represents the mixed phase of negatively charged normal quark matter and
positively charged 2SC matter.}
\end{figure}

Let us start by giving a brief introduction into the general method of
constructing mixed phases by imposing the Gibbs conditions of equilibrium
\cite{glen92,Weber}. From the physical point of view, the Gibbs
conditions enforce the mechanical as well as chemical equilibrium between
different components of a mixed phase. This is achieved by requiring that
the pressure of different components inside the mixed phase are equal,
and that the chemical potentials ($\mu$ and $\mu_{e}$) are the same
across the whole mixed phase. For example, in relation to the mixed phase
of normal and 2SC quark matter, these conditions read
\begin{eqnarray}
P^{(NQ)}(\mu,\mu_{e}) &=& P^{(2SC)}(\mu,\mu_{e}), 
\label{P=P}\\ 
\mu &=& \mu^{(NQ)}=\mu^{(2SC)}, 
\label{mu=mu}\\ 
\mu_{e} &=& \mu^{(NQ)}_{e}=\mu^{(2SC)}_{e}.  
\label{mue=mue}
\end{eqnarray} 
It is easy to visualize these conditions by plotting the pressure as a
function of chemical potentials ($\mu$ and $\mu_{e}$) for both components
of the mixed phase. This is shown in Fig. \ref{fig-front-quark}. As
should be clear, the above Gibbs conditions are automatically satisfied
along the intersection line of two pressure surfaces (dark solid line in
Fig. \ref{fig-front-quark}).

Different components of the mixed phase occupy different volumes of
space. To describe this quantitatively, we introduce the volume fraction
of normal quark matter as follows: $\chi^{NQ}_{2SC}\equiv V_{NQ}/V$
(notation $\chi^{A}_{B}$ means ``volume fraction of phase A in a mixture
with phase B"). Then, the volume fraction of the 2SC phase is given by
$\chi^{2SC}_{NQ}=(1-\chi^{NQ}_{2SC})$. From the definition, it is clear
that $0\leq \chi^{NQ}_{2SC} \leq 1$.

The average electric charge density of the mixed phase is determined by
the charge densities of its components taken in the proportion of the
corresponding volume fractions. Thus,
\begin{equation}
n^{(MP)}_{e} = \chi^{NQ}_{2SC} n^{(NQ)}_{e}(\mu,\mu_e) 
+(1-\chi^{NQ}_{2SC}) n^{(2SC)}_{e}(\mu,\mu_e).
\end{equation}
If the charge densities of the two components have opposite signs, one
can impose the global charge neutrality condition, $n^{(MP)}_{e}=0$.
Otherwise, a neutral mixed phase could not exist. In the case of 
quark matter, the charge density of the normal quark phase is negative,
while the charge density of the 2SC phase is positive along the line of
the Gibbs construction (dark solid line in Fig. \ref{fig-front-quark}).
Therefore, a neutral mixed phase exists. The volume fractions of its
components are
\begin{eqnarray}
\chi^{NQ}_{2SC} &=& \frac{n^{(2SC)}_{e}}{n^{(2SC)}_{e}-n^{(NQ)}_{e}}, \\
\chi^{2SC}_{NQ} &\equiv& 1-\chi^{NQ}_{2SC}=
\frac{n^{(NQ)}_{e}}{n^{(NQ)}_{e}-n^{(2SC)}_{e}}.
\end{eqnarray}

After the volume fractions have been determined from the condition of the
global charge neutrality, we could also calculate the energy density of
the corresponding mixed phase,
\begin{equation}
\varepsilon^{(MP)} = \chi^{NQ}_{2SC} \varepsilon^{(NQ)}(\mu,\mu_e)
+(1-\chi^{NQ}_{2SC}) \varepsilon^{(2SC)}(\mu,\mu_e).
\end{equation}
This is essentially all that we need in order to construct the equation of
state of the mixed phase. 

So far, we were neglecting the effects of the Coulomb forces and the
surface tension between different components of the mixed phase. In a real
system, however, these might be important. In particular, the balance
between the Coulomb forces and the surface tension determines the
geometries of different components inside the mixed phase. 

In our case, nearly equal volume fractions of the two quark phases are
likely to form alternating layers (slabs) of matter. The energy cost per
unit volume to produce such layers scales as $\sigma^{2/3}
(n_{e}^{(2SC)}-n_{e}^{(NQ)})^{2/3}$ where $\sigma$ is the surface tension
\cite{geometry}. Therefore, the quark mixed phase is a favorable phase of
matter only if the surface tension is not too large. Our simple estimates
show that $\sigma_{max} \alt 20$ MeV/fm$^{2}$. However, even for slightly
larger values, $20 \alt \sigma \alt 50$ MeV/fm$^{2}$, the mixed phase is
still possible, but its first appearance would occur at larger densities,
$3\rho_0 \alt \rho_B \alt 5\rho_0$. The value of the maximum surface
tension obtained here is comparable to the estimate in the case of the
hadronic-CFL mixed phase obtained in Ref. \cite{interface}. The thickness
of the layers scales as $\sigma^{1/3} (n_{e}^{(2SC)}-n_{e}^{(NQ)})^{-2/3}$
\cite{geometry}, and its typical value is of order $10$ fm in the quark
mixed phase. This is similar to the estimates in various hadron-quark and
hadron-hadron mixed phases \cite{geometry,interface}. While the actual
value of the surface tension in quark matter is not known, in this study
we assume that it is not very large. Otherwise, the homogeneous gapless
2SC phase should be the most favorable phase of nonstrange quark matter
\cite{HS}.

Under the assumptions of this paper, the mixed phase of normal and 2SC
quark matter is the most favorable neutral phase of matter in the model
at hand. This should be clear from observing the pressure surfaces in
Figs. \ref{fig-back} and \ref{fig-front-quark}. For a given value of the
baryon chemical potential $\mu=\mu_{B}/3$, the mixed phase is more
favorable than the gapless 2SC phase, while the gapless 2SC phase is more
favorable than the neutral normal quark phase.

The validity of the quark model is limited when the baryon density
decreases. From the physical point of view, at some point quark matter
should become confined and the NJL model should fail to reproduce the
correct equation of state of low density baryonic matter. Therefore, at
low densities, it is appropriate to use a hadronic description of matter.
This is discussed in the next section.

\section{Low density region and quark-hadron phase transition}
\label{low-density}

At densities around normal nuclear matter density $\rho_0\approx 0.15$
fm$^{-3}$, the description of baryonic matter in terms of quarks could
hardly be adequate. At such low densities, quarks are confined inside
hadrons. Thus, it is more natural to use an effective hadronic model.

In the literature, there exist many hadronic models that have been
studied in detail. In particular, the Walecka model
\cite{sero86,walebuch} and its nonlinear extensions \cite{sch96a,Pal} are
well known. These models have been quite successful and widely used for
the description of finite nuclei, hadronic and neutron star matter. One
could also use some hadronic versions of the NJL model
\cite{njl61a,reber96a,mishu99}, as well as many others \cite{bogu,fpw}.

In the following we use a QCD-motivated hadronic chiral SU(3)$_L \times
$SU(3)$_R$ model as an effective theory of strong interactions to
describe the low density regime of the baryonic matter
\cite{papa98,papa99,han2000}. This model was found to describe reasonably
well the hadronic masses of various SU(3) multiplets, finite nuclei,
hypernuclei, excited nuclear matter and neutron star properties
\cite{papa98,papa99,han2000}. The basic features of the chiral model 
are:
(i) Lagrangian of the model is constructed in accordance with the
nonlinear realization of the chiral SU(3)$_L \times $SU(3)$_R$ symmetry; 
(ii) heavy baryons and mesons get their masses as a result of
spontaneous symmetry breaking;
(iii) the masses of pseudoscalar mesons (pseudo-Nambu-Goldstone
bosons) result from an explicit symmetry breaking;
(iv) a QCD-motivated field that describes the gluon condensate 
(dilaton field) is introduced.

It is expected that the phase transition between the hadronic phase and
the normal quark phase is a first order phase transition at zero
temperature and finite baryon chemical potential. Then, the hadronic and
quark phases could co-exist in a mixed phase \cite{glen92,Weber}. This
mixed phase should satisfy the Gibbs conditions of equilibrium which 
are similar to those discussed in the previous section, see Eqs.
(\ref{P=P})--(\ref{mue=mue}). The total energy density in the
hadron-quark mixed phase is given by
\begin{eqnarray}
\varepsilon^{(MP)} = \chi^{NQ}_{H} \epsilon^{(NQ)}(\mu,\mu_e)
+(1-\chi^{NQ}_{H}) \epsilon^{(H)}(\mu,\mu_e) ,
\label{eq:epsilon_MP}
\end{eqnarray}
where $\chi^{NQ}_{H}$ denotes the volume fraction of normal quark 
matter inside a mixture with hadronic matter.

\begin{figure}
\includegraphics[bbllx=5,bblly=45,bburx=575,bbury=590,width=8cm]
{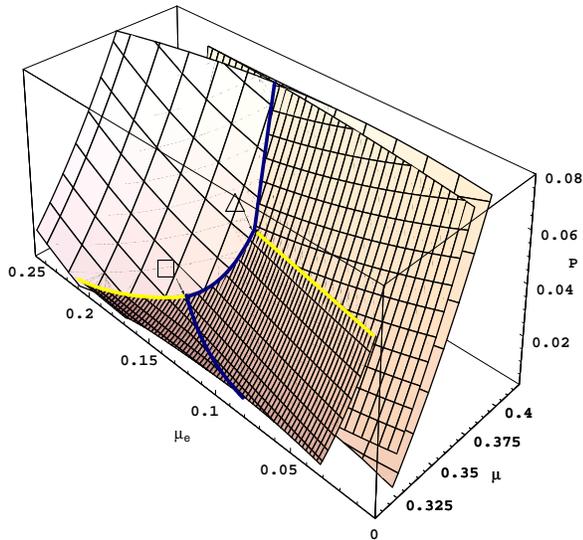}
\caption{\label{fig:3D_HadQu}
Pressure as a function of $\mu\equiv\mu_B/3$ and $\mu_e$ for the
hadronic phase (at the bottom), for the two-flavor color superconducting   
phase (on the right in front) and the normal phase of quark matter (on
the left, and back on the right). The dark thick line follows the
neutrality line in hadronic matter, and two mixed phases: (i) the
mixed phase of hadronic and normal quark matter; and (ii) the mixed
phase of normal and color superconducting quark matter.}
\end{figure}

To visualize the Gibbs construction of the hadron-quark mixed phase, we
plot the hadronic surface of the pressure along with the quark surfaces
discussed in the previous section. Thus, Fig. \ref{fig-front-quark} is
replaced by Fig.~\ref{fig:3D_HadQu}. The new figure shows the surfaces of
the pressure of the pure hadronic and quark phases as a function of their
chemical potentials $\mu$ and $\mu_e$. The intersection lines of different
surfaces indicate all potentially viable mixed phase constructions.
Although the Gibbs conditions are satisfied along all these lines, not all
of them could produce globally neutral phases (e.g., there are no neutral
constructions along the light shaded solid lines in
Fig.~\ref{fig:3D_HadQu}).

The dark solid line gives the complete, most favorable construction of
globally neutral matter in general $\beta$-equilibrium. This line consists
of three pieces. The lowest part of the curve (up to the point denoted by
$\square$ and $P\alt 10$ MeV$/$fm$^{3}$) corresponds to the pure confined
hadronic phase. Within this region matter is mostly composed of neutrons
with little fractions of protons and electrons to realize the charge
neutrality and $\beta$-equilibrium. The hyperonic particles ($\Lambda,
\Sigma$ and $\Xi$) are not present in this lowest density region. Such
particles would appear in the hadronic phase at considerably higher values
of pressure and density.

The mixed phase of hadronic and normal quark matter starts at the baryonic
density $\rho_B\approx 1.49 \rho_0$ which corresponds to the
$\square$-point in Fig.~\ref{fig:3D_HadQu}. At this point the first
bubbles of deconfined quark matter appear in the system. At the beginning
of this hadron-quark mixed phase, the deconfined bubbles are small but
highly negatively charged, whereas hadronic matter, in which the bubbles
are embedded, is slightly positively charged. The global charge neutrality
condition reads
\begin{equation} 
n^{(MP)}_{e} \equiv \chi^{NQ}_{H} n^{(NQ)}_{e}(\mu,\mu_e)
+(1-\chi^{NQ}_{H}) n^{(H)}_{e}(\mu,\mu_e) =0.
\label{eq:q_MP} 
\end{equation}
where $n^{(H)}_{e}$ and $n^{(NQ)}_{e}$ are the charge densities of
hadro\-nic and normal quark matter, respectively. This condition should be
satisfied at each point along the middle part of the dark solid line
(i.e., between the points denoted by $\square$ and $\bigtriangleup$).
With increasing density (from about $1.49 \rho_0$ up to $2.56 \rho_0$), 
the volume fraction of hadronic matter decreases (down to about
$0.59$), while the fraction of normal quark matter increases (up to
about $0.41$). 

It is quite remarkable that the hadron-quark mixed phase does not reach a
point where the fraction of hadronic matter would vanish completely. This
is in contrast to other examples of hadron-quark Gibbs constructions in
the literature \cite{glen92,schert99,mishu2003}. Instead, at baryon
density about $2.56 \rho_0$ (a point denoted by $\bigtriangleup$ in
Fig.~\ref{fig:3D_HadQu}), the mixed phase is replaced by another mixed
phase which is made of the normal and 2SC quark components. At this
point, positively charged hadronic matter will be suddenly converted into
positively charged color superconducting quark matter in the 2SC phase.
As a result of this rearrangement, the values of the baryon density and
the energy density experience small jumps: the baryon density changes
from about $2.56 \rho_0$ to $2.75 \rho_0$ and the energy density changes
from $378$ MeV$/$fm$^{3}$ to $415$ MeV$/$fm$^{3}$. Right after the
transition, the volume fractions of the 2SC and normal quark phases are
$0.53$ and $0.47$, respectively.

The mixed phase of normal and 2SC quark matter above the point
$\bigtriangleup$ in Fig.~\ref{fig:3D_HadQu} will remain the most
favorable globally neutral phase of baryonic matter in our model. The
volume fractions of the two components of this phase stay nearly constant
with increasing density. This unusual property of the mixed phase
originate from the simple fact that 2SC quark matter has a strong
preference to remain positively charged. In principle, such a mixed phase
could possibly be replaced by a neutral 2SC+s or CFL phase (or, possibly,
even by some new mixed phase) at sufficiently high densities when the
strange quarks appear in the system. However, it may also happen that the
stars with high enough densities become gravitationally unstable long
before the strange quarks get a chance to appear.

\begin{figure}
\includegraphics[bbllx=75,bblly=240,bburx=545,bbury=730,width=8cm]
{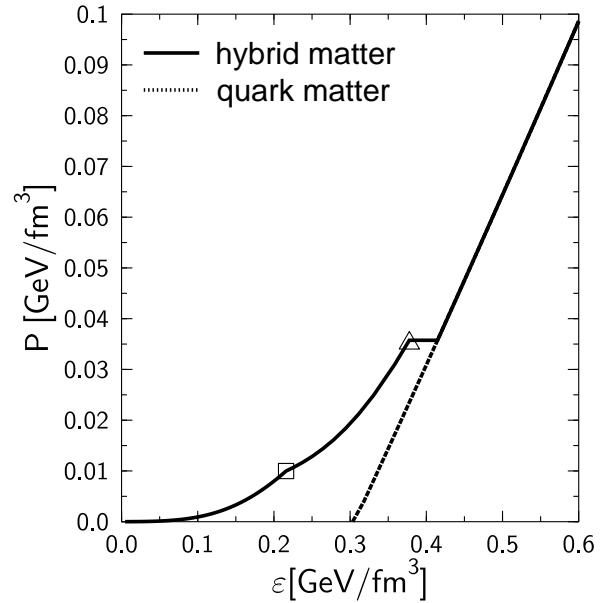} 
\caption{\label{fig:eos} 
The equation of state for globally neutral hybrid matter (solid line)  
and globally neutral quark matter (dashed line). The points of the
beginning of the two mixed phases are denoted by $\square$ and
$\bigtriangleup$.}
\end{figure}

The equations of state for quark and hybrid matter are shown in
Fig.~\ref{fig:eos}. The first equation of state corresponds to globally
neutral quark matter which is a mixture of the normal quark and 2SC
phases. This was discussed in Sec. \ref{quark-Gibbs} in detail, see
Fig.~\ref{fig-front-quark}. As for the equation of state of hybrid
matter, it is constructed out of the equation of state of neutral
hadronic matter and two Gibbs constructions in accordance with
Fig.~\ref{fig:3D_HadQu}. As before, the points that indicate the
beginning of two different mixed phases are denoted by $\square$ and
$\bigtriangleup$ in Fig.~\ref{fig:eos}.

\section{Star properties}
\label{star-properties}

It is evident from Fig.~\ref{fig:eos} that no minimum appears in the
equation of state of hybrid matter. Therefore, the corresponding
compact stars can only be bound by the gravitational force. Because of a
strong gravitational field inside and around such compact stars, we need
to describe the gravitational field within the framework of general
relativity as a curvature of spacetime. By knowing the behavior of the
pressure and the energy density of matter inside the star, one can obtain
the corresponding spacetime geometry by solving the Einstein equation,
\begin{equation}
R_{\mu\nu} - \frac{1}{2} g_{\mu\nu} R = 8 \pi \kappa T_{\mu\nu},
\label{eq:Einsteineq}
\end{equation}
where $R_{\mu\nu}$ is the Ricci tensor, $g_{\mu\nu}$ is the spacetime
metric and $T_{\mu\nu}$ is the energy-momentum tensor of matter
inside the compact star.

\begin{figure}
\includegraphics[bbllx=75,bblly=240,bburx=545,bbury=730,width=8cm]
{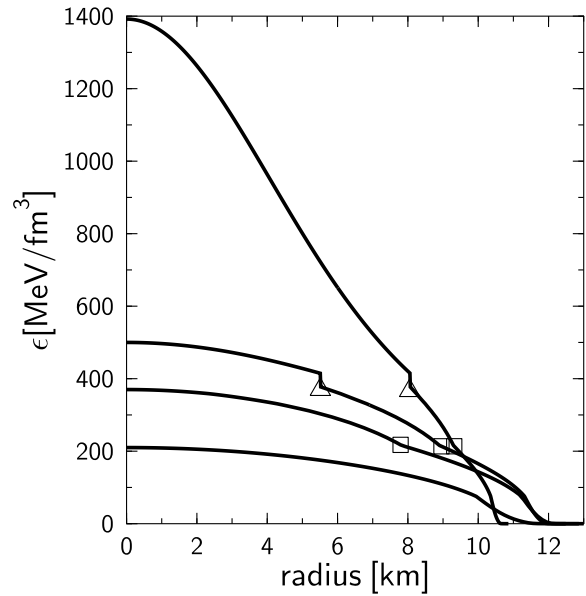}
\caption{\label{fig:er}
Energy density profiles for hybrid stars. The locations of the interface
between the two types of mixed phases are denoted by $\bigtriangleup$,   
while the locations of the boundary between the pure hadronic phase and
the hadron-quark mixed phase are denoted by $\square$.}
\end{figure}

The metric inside a non-rotating spherically symmetric star with the
energy-momentum tensor $T_{\mu\nu}$ of an ideal fluid \cite{MTW} can be
determined by solving the Tolman-Oppenheimer-Volkoff (TOV) equations
\cite{Tol} that follow from the above Einstein equation. The metric is
given by the following ansatz: 
\begin{equation} 
g_{\mu\nu}= \mbox{diag} \left( e^{2\nu(r)},
-\left(1-\frac{2m(r)}{r}\right)^{-1}, 
-r^2, -r^2{\sin}^2\theta \right), 
\end{equation} 
and the explicit form of the TOV equations read
\begin{eqnarray} 
m(r) &=& 4 \pi \int_{0}^{r} \epsilon(r_{1}) r_{1}^2 dr_{1}, \\
\frac{d\nu}{dr} &=& \frac{m(r) + 4 \pi r^3 P(r)}
{r\left[r - 2 m(r)\right]}, \\
\frac{dP}{dr} &=& -\left[P(r) + \epsilon(r)\right] \frac{d\nu}{dr}. 
\label{eq:8} 
\end{eqnarray} 
(To simplify notation, we use units with $\kappa=c=1$.) The functions
$P(r)$ and $\epsilon(r)$ are the pressure and energy density at radius
$r$ inside the star. By definition, $R$ is the radius of the star while
$M\equiv m(R)$ is the star mass. For a given equation of state
$P(\epsilon)$ and a fixed central energy density $\epsilon_c \equiv
\epsilon(r=0)$ one can numerically integrate the above set of equations
from the centre of the star up to its surface ($r=R$) where the pressure
is zero, i.e., $P(R)=0$, and 
\begin{equation} 
\nu(r) \equiv \frac{1}{2}\ln \left( 1- \frac{2 M}{r}\right) ,
\quad \mbox{for} \quad r\geq R. 
\end{equation} 
The energy density profiles for hybrid stars with different central
energy densities are displayed in Fig.~\ref{fig:er}. The star with the
lowest central energy density in Fig.~\ref{fig:er}, $\epsilon_c=210$
MeV$/$fm$^{3}$, is composed of pure confined hadronic matter, mainly
neutrons, surrounded by a thin compact star crust consisting of leptons
and nuclei. To get the equation of state for the crust, we use the
results of Ref. \cite{bay71a} for $\rho_B<0.001$ fm$^{-3}$ and the
results of Ref. \cite{nege73a} for $0.001$ fm$^{-3}$ $< \rho_B < 0.08$
fm$^{-3}$.

The next energy density profile in Fig. \ref{fig:er} corresponds to the
central energy density $\epsilon_c=370$ MeV$/$fm$^{3}$. We see that the
corresponding star already has a rather large core (the radius is about
$8$ km) consisting of a mixture of hadronic and normal quark
matter. The core of the star is surrounded by a layer of hadronic matter
and a crust.

As we saw in the previous section, there is no globally neutral baryonic
matter that would produce the energy densities in the window between
$378$ MeV$/$fm$^{3}$ and $415$ MeV$/$fm$^{3}$. Therefore, there are no
stars that could have the central energy densities in this window either.
At $\epsilon_c>415$ MeV$/$fm$^{3}$, a quark core (made of the mixed phase
of normal quark and 2SC matter) forms at the center of the star. Two
examples of the corresponding energy density profiles are also shown in
Fig. \ref{fig:er}. The star with the central energy density
$\epsilon_c=500$ MeV$/$fm$^{3}$ contains a quark phase core with radius
about $6$ km. This core is separated from the layer of the hadron-quark
mixed phase by a sharp interface (the corresponding point is denoted by
$\bigtriangleup$ in Fig. \ref{fig:er}). The most extreme stable star
within this model (the star with the largest possible mass), has the
central energy density $\epsilon_c=1392$ MeV$/$fm$^{3}$. This star is
mainly composed of the quark core surrounded by a relatively thin layer
of the hadron-quark mixed phase, as well as the pure hadronic phase and
the crust on the outside.

\begin{figure}
\includegraphics[bbllx=75,bblly=240,bburx=545,bbury=730,width=8cm]  
{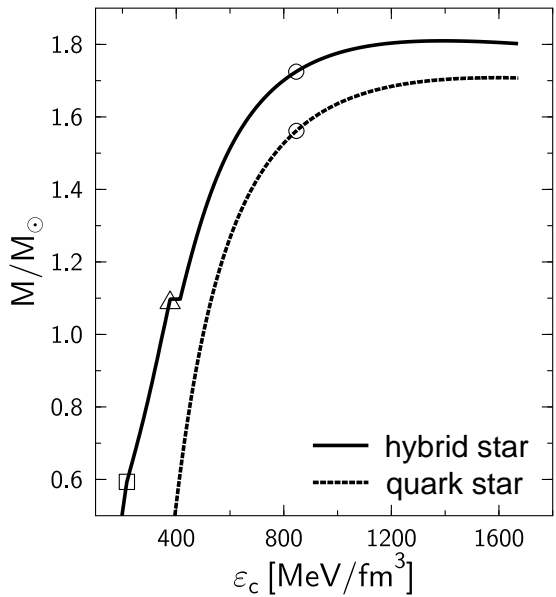}
\caption{\label{fig:md}
The gravitational mass $M$ versus the central energy density for hybrid
stars (solid line) and quark stars (dashed line). The corresponding
equations of state are shown in Fig. \ref{fig:eos}.  The stars heavier
than the ``$\bigcirc$-star" have central baryon densities larger than
$5\rho_0$. Such stars may have seeds of strange quark matter at their
cores.}
\end{figure}

The dependence of the gravitational mass $M$ as a function of the central
energy density $\epsilon_c$ for hybrid and quark stars are displayed in
Fig.~\ref{fig:md}. The first appearance of the two mixed phases inside
hybrid stars are marked with the symbols $\square$ and $\bigtriangleup$.
The maximum mass hybrid star has the following properties: $M_{\rm
max}=1.81 M_\odot$, $\epsilon_c=1392$ MeV$/$fm$^{3}$, $\rho_c=7.58
\rho_0$ and $R=10.86$ km. Hybrid stars with higher masses or central
densities would collapse to black holes because of gravitational
instability.

\begin{figure}
\includegraphics[bbllx=75,bblly=240,bburx=545,bbury=730,width=8cm]  
{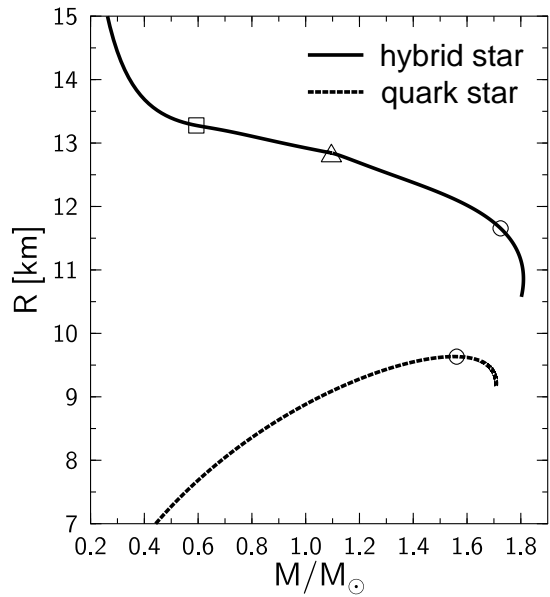}
\caption{\label{fig:mr}
The mass-radius relations for hybrid stars (solid line) and quark stars
(dashed line).}
\end{figure}

The mass-radius relations for hybrid and quark stars are shown in
Fig.~\ref{fig:mr}. As one would expect, the pure quark stars composed of
the mixed phase constructed in Sec. \ref{quark-Gibbs} have much smaller
radii and the value of their maximum mass is slightly smaller, see
Fig.~\ref{fig:mr}. The difference between hybrid and pure quark stars is
mostly due to the low density part of the equation of state. This is also
evident from the qualitative difference in the dependence of the radius
as a function of mass for the hybrid and quark stars with low masses. The
corresponding hybrid stars are large because of a sizable low density
hadronic layer, while the quark stars are small because that have no such
layers.

Our results for pure quark stars are comparable to those in Refs.
\cite{Blaschke_2sc,Ruester,super-dense,compact_CFL}. Also, the maximum
masses and the corresponding radii of the hybrid stars obtained here are
similar to those of the strange hybrid stars of Ref. \cite{super-dense},
assuming that the strange quark mass is not very small ($m_s\agt 300$ MeV)
and the superconducting gap is not too large ($\Delta \alt 50$ MeV). At
smaller values of the strange quark mass and/or larger values of the
superconducting gap, the strange hybrid stars tend to have smaller
maximum masses and smaller radii \cite{super-dense}.

In this paper we use a two-flavor version of the NJL model to describe the
deconfined phase. However, the three-flavor extension of the NJL model in
Ref. \cite{neutral_steiner} suggests that strange quarks might be present
in matter above a critical density of about $5\rho_0$. In accordance with
this, we have marked the points of the expected appearance of strange
quarks in the center of the stars with the symbol $\bigcirc$ in
Figs.~\ref{fig:md} and \ref{fig:mr}. For a more realistic picture of
hybrid and quark stars with masses and central energy densities larger
than those of the ``$\bigcirc$-star", strange quarks may need to be
included in the description. The corresponding study of compact stars is
left for future work. It is worth to emphasize, however, that the
properties of the hybrid stars lighter than the ``$\bigcirc$-star"  in
Figs.~\ref{fig:md} and \ref{fig:mr}, would remain unchanged.

\section{Summary and outlook}
\label{summary}

In this paper, we constructed a realistic equation of state of nonstrange
baryonic matter that is globally neutral and satisfies the condition of
$\beta$-equilibrium. This equation of state might be valid up to
densities of about $5\rho_0$, or even up to densities of about $8\rho_0$
in the most optimistic scenario. In our construction, matter at low
density ($\rho_B\alt 1.49 \rho_0$) is mostly made of neutrons with traces
of protons and electrons. At intermediate densities ($1.49 \rho_0 \alt
\rho_B \alt 2.56 \rho_0$), homogeneous hadronic matter is replaced by the
mixed phase of positively charged hadronic matter and deconfined bubbles
of negatively charged normal quark matter. The volume fraction of normal
quark matter gradually grows with increasing density. Before the volume
fractions of two phases become equal, the mixed phase undergoes a
rearrangement in which the hadronic component of matter turns into the
two-flavor color superconductor.  At higher densities ($\rho_B \agt 2.75
\rho_0$), only the quark mixed phase exists. This latter is composed of
about equal fractions of normal and 2SC quark matter.

Previously, it was argued that 2SC quark matter could not appear in
compact stars when the charge neutrality condition is imposed
\cite{absence2sc}. The main reason for this is a strong preference of
2SC quark matter to remain positively charged. Therefore, one may
conclude that no color superconducting phases can appear inside compact
stars if the baryon density is not large enough for the strange quarks to
condense and participate in the formation of the neutral CFL phase. The
present study reinstalls the status of 2SC quark matter as the most
promising color superconducting phase in the central regions of compact
stars.

Our study shows that positively charged 2SC quark matter appears
naturally in a quark mixed phase at densities around $3\rho_0$. The other
component of the globally neutral mixed phase is negatively charged
normal quark matter. The corresponding construction turns out to be very
stable. In particular, we observe that the volume fractions of the two
quark components remain approximately the same with changing the baryon
density. From the physical point of view, this rigidity of the mixed
phase is connected with the nature of 2SC quark matter which tends to
remain positively charged.

To the best of our knowledge, hybrid baryonic matter in this paper gives
the first and the only example of a combination of two different Gibbs
constructions that replace each other in the same system with changing the
density. It should be clear that this special construction results from
the existence of the triple point where hadronic matter coexists with two
different quark phases. It is fare to mention that some speculations about
the possibility of another type of a triple point were expressed in Ref.
\cite{triple}. In our construction, at the triple point, the hadronic
component of the first mixed phase is replaced by the 2SC quark component
of the other. This could be interpreted as a first order phase transition
that happens only in one of the components of the mixed phase. The volume
fraction of the inert component, i.e., normal quark matter, changes by a
small jump during this transition. This change accounts for the difference
of the charge densities of hadronic and 2SC quark matter.

By making use of the equation of state of hybrid matter, we construct
the corresponding non-rotating compact stars. We find that the largest
mass hybrid star has the radius $10.86$ km, the mass $1.81 M_{\odot}$ and
the central baryon density $7.58 \rho_0$. This star has a large ($8$ km)
quark core, and a relatively thin outer layers of hadronic matter and a
crust. It also appears that the quark cores are quite large even for
stars with relatively low central energy densities. Clearly, this is the
reflection of the fact that the mixed quark phase starts to develop at
rather low densities, $\rho_B\approx 2.75\rho_0$, in the NJL model
used. 

The quark core in a hybrid star is separated from the lower density
hadron-quark layer by a sharp interface where the baryon density and the
energy density change by a jump. In a real star, this interface should be
naturally smoothed over the distances comparable to the physical sizes of
different matter components in the mixed phase. Our estimates show that a
typical thickness of the layers of different quark phases is about $10$
fm. This should also be comparable to the physical sizes of components in
the hadron-quark mixed phase around the interface
\cite{geometry,interface}. This value sets the scale of the actual size of
the interface. A more detailed study of the interface between the mixed
phases is outside the scope of this paper.

In our analysis, we used a rather simple quark model and a specific
hadronic model. This may suggest that most of our conclusions in this
paper are not quite general. We would like to argue, however, that is not
the case. In fact, the main results should remain qualitatively correct
and model independent, because they are based on a simple and clear
physical picture. For example, the fact that the most favorable 2SC quark
phase appears to be positively charged is related to the nature of the
phase itself rather than to any specific properties of the model. It is
this fact that was used previously to justify that the CFL phase is more
favorable than the homogeneous 2SC (or 2SC+s) phase in compact stars.
Here we turn the same arguments around, and find that the mixed phase of
normal and 2SC quark matter is the most favorable nonstrange globally
neutral quark matter. Indeed, when converting all quark matter into a
color superconductor is impossible, the next best possibility is to
convert at least a fraction of it.

It is fare to mention that, in this paper, we made a rather conservative
assumption that the strange quark is very heavy and it does not appear in
quark matter even at rather high densities. In particular, the highest
density reached in our model of the hybrid star without causing a
gravitational instability is $7.58 \rho_0$. We appreciate that our
assumption may not be justified at such large densities. In fact, model
calculations of Ref. \cite{neutral_steiner} suggest that the strange
quarks start to appear in quark matter at baryon densities around
$5\rho_0$ (this corresponds to $\mu=\mu_{B}/3\approx 450$ MeV). If we
accept this, the hybrid stars which are heavier than ``$\bigcirc$-stars"
in Figs. \ref{fig:md} and \ref{fig:mr} should contain some strange matter
at their most central regions. Apparently, the appearance of the
strangeness would result in reducing the maximum value of the mass of the
hybrid star. As one can see from Fig. \ref{fig:mr}, however, there is no
much room for a big change of the maximum mass. Of course, adding
strangeness may also open the possibility of a new generation of stars
with considerably smaller radii \cite{super-dense,twins}.

In the future, it would be interesting to study the physical properties of
the hybrid stars that have not been addressed in this paper. For example,
it would be interesting to study the neutrino emissivity and the mean free
path of neutrinos inside quark cores of such stars. The corresponding
results would be crucial for understanding the cooling mechanism of the
hybrid stars. Also, it would be interesting to address the magnetic and
various transport properties. The nature of the quark mixed phase suggests
that a flux of the magnetic field should penetrate through the quark core.
However, it might also be interesting to see how the flux is distributed
inside the mixed phase, and how this may affect the star properties.
Finally, it would be useful to generalize the current study to the case of
rotating stars.

\begin{acknowledgments}
The authors thank D.H.~Rischke and J.~Schaffner-Bielich for useful
discussions, as well as M.~Alford and K.~Rajagopal for comments on the
first version of the manuscript.  M. Hanauske acknowledges the financial
support from the Hessische Ministerium f\"ur Wissenschaft und Kunst. M.
Huang acknowledges the financial support from Bundesministerium f\"{u}r
Bildung und Forschung (BMBF), the Alexander von Humboldt-Foundation, and
the NSFC under Grant Nos.  10105005, 10135030. The work of I. Shovkovy was
supported by Gesellschaft f\"{u}r Schwerionenforschung (GSI) and by
Bundesministerium f\"{u}r Bildung und Forschung (BMBF).
\end{acknowledgments}

\end{document}